\documentclass[12pt,onecolumn,letterpaper]{IEEEtran}
\usepackage[T1]{fontenc}
\usepackage[latin1]{inputenc}
\usepackage{color}
\usepackage{psfrag}
\usepackage[dvips]{graphicx}
\usepackage{enumerate}
\usepackage{rotating}
\usepackage{srcltx}
\usepackage[noadjust]{cite}
\usepackage[cmex10]{amsmath}
\usepackage{amsfonts}
\usepackage{amssymb}
\usepackage{ntheorem}
\usepackage{dsfont}    
\usepackage{mathtools} 
\usepackage{algorithmic}
\usepackage{array}
\usepackage[tight,footnotesize]{subfigure}
\newcommand{\Prob}{{\mathbb{P}}}

\newcommand{\bern}{\mathop{\mathrm{Bern}}\nolimits}
\theoremstyle{plain}
\theoremheaderfont{\itshape}\theorembodyfont{\itshape}
\theoremseparator{.}
\newtheorem{theorem}{Theorem}
\newtheorem*{corollary*}{Corollary}

\theoremheaderfont{\itshape}\theorembodyfont{\upshape}
\theoremseparator{.}
\newtheorem{definition}{Definition}
\newtheorem{remark}{Remark}
\newtheorem*{example*}{Example}

\long\def\symbolfootnote[#1]#2{\begingroup%
\def\thefootnote{\fnsymbol{footnote}}\footnote[#1]{#2}\endgroup} 
\begin{document}
%
\title{On Secure Computation Over the Binary Modulo-2 Adder Multiple-Access Wiretap Channel\vspace{10pt}}
\author{\IEEEauthorblockN{Mario~Goldenbaum\IEEEauthorrefmark{1}, Holger~Boche\IEEEauthorrefmark{2}, and H. Vincent~Poor\IEEEauthorrefmark{1}\\ \vspace{7pt}}
{\normalsize\IEEEauthorblockA{\IEEEauthorrefmark{1}Department of Electrical Engineering, Princeton University\\}
\IEEEauthorblockA{\IEEEauthorrefmark{2}Chair of Theoretical Information Technology, Technical University of Munich}}}
\bstctlcite{mario_goldenbaum:BSTcontrol}
%
%
\maketitle
\symbolfootnote[0]{This work was supported in part by the German Research Foundation (DFG) under grant GO 2669/1-1 and by the U. S. National Science Foundation under Grant CMMI-1435778.}
%
%
%
%
\begin{abstract}
	In this paper, the problem of securely computing a function over the binary modulo-2 adder multiple-access wiretap channel is considered. The problem involves a legitimate receiver that wishes to reliably and efficiently compute a function of distributed binary sources while an eavesdropper has to be kept ignorant of them. In order to characterize the corresponding fundamental limit, the notion of \emph{secrecy computation-capacity} is introduced. Although determining the secrecy computation-capacity is challenging for arbitrary functions, it surprisingly turns out that if the function perfectly matches the algebraic structure of the channel and the joint source distribution fulfills certain conditions, the secrecy computation-capacity equals the computation capacity, which is the supremum of all achievable computation rates without secrecy constraints. Unlike the case of securely transmitting messages, no additional randomness is needed at the encoders nor does the legitimate receiver need any advantage over the eavesdropper. The results therefore show that the problem of securely computing a function over a multiple-access wiretap channel may significantly differ from the one of securely communicating messages.
\end{abstract}%
%
%
\section{Introduction} \label{sec:introduction}%
In their seminal work \cite{Nazer:Gastpar:07b}, Nazer and Gastpar lay the information-theoretic foundation of distributed computation over unreliable channels. The big difference between this approach and the standard theory dealing with reliable message transfer is that, in \cite{Nazer:Gastpar:07b}, the intended receiver decodes function values immediately from the channel output. In other words, the receiver does not care about individual messages and penalizes itself only when the function is incorrectly decoded. 

In this regard, Nazer and Gastpar show that in many cases, the performance gain over separation-based computation strategies is proportional to the number of source terminals. In a separation-based strategy, the receiver first reliably decodes all individual messages and subsequently computes the sought function value. It is remarkable that the gains over separation-based strategies stem from a match between the desired function and the algebraic structure of the channel. Since the publication of \cite{Nazer:Gastpar:07b}, the results and ideas have been extended in many different ways \cite{Soundararajan:Vishwanath:2012,Goldenbaum:Stanczak:13a,Karamchandani:Niesen:Diggavi:13,Goldenbaum:Boche:Stanczak:14,Wang:Jeon:Gastpar:13b}.

Due to the trend towards large-scale decentralized networks consisting of many mutually distrusting terminals, \emph{security and integrity} of computation results are of high priority in order to guarantee trustworthy operation. In this work, we therefore make a first attempt to extend the concept of computation coding \cite{Nazer:Gastpar:07b} by taking information theoretic security aspects into account. In particular, we consider the problem of computing a function over the binary modulo-2 adder multiple-access wiretap channel (MAWC). The problem involves a legitimate receiver that wishes to reliably compute a function of distributed binary sources in the presence of an eavesdropper. To characterize the corresponding fundamental limit, we introduce the notion of \emph{secrecy computation-capacity}. Although determining the secrecy computation-capacity for arbitrary functions is challenging, it turns out that if the function perfectly matches the algebraic structure of the modulo-2 adder MAWC and the joint source distribution fulfills certain conditions, the secrecy computation-capacity \emph{equals} the computation capacity. Thus, the algebraic structure of the channel not only helps to efficiently compute the desired function but also to protect the transmitted source sequences against eavesdropping. It is noteworthy that, to achieve this, the source terminals \emph{do not need} any additional source of randomness nor does the legitimate receiver need any advantage over the eavesdropper. This is in contrast to standard physical-layer security results.
%
%
\subsection{Related Work} \label{sec:related_work}%
%
Considering secure distributed computation, also known as secure multi-party computation, from an information theoretic (i.e., Shannon) perspective is still in its infancy. To the best of the authors' knowledge there exist only some very recent results. For instance, Tyagi et al. introduce a new Shannon theoretic multiuser source model in \cite{Tyagi:Narayan:Gupta:11} and \cite{Tyagi:13} and characterize when a function is securely computable. In this context, they provide necessary and sufficient conditions for the existence of protocols that achieve this. 

Within the standard secure multi-party computation model of \cite{Yao:82}, Lee and Abbe determine in \cite{Lee:Abbe:14} the least amount of randomness needed for securely computing a given function. This provides a novel notion of the complexity of a function for its secure computation. In the second part of that paper, the considerations are extended to a probabilistic source model for which the decoding error probability is required to vanish asymptotically in the block length.

In \cite{Data:Dey:Mishra:Prabhakaran:14}, Data et al. take a distributed source coding approach to the problem of securely computing the modulo-2 sum of two distributed binary sources. Similarly to \cite{Lee:Abbe:14}, they assume the data to be drawn from some joint memoryless source and derive bounds on the amount of randomness and communication needed to asymptotically achieve secrecy. In \cite{Data:Prabhakaran:15}, the results are extended to arbitrary functions.

All these works are through the lens of source coding, which means that the communication between terminals is assumed to take place over noiseless channels. In this paper, we therefore choose a \emph{joint source-channel coding} perspective.
%
%
\subsection{Paper Organization} \label{sec:contributions}%
%
This paper is organized as follows. Section~\ref{sec:system_model} introduces the binary modulo-2 adder MAWC and provides the problem statement. In order to obtain some insight, in Section~\ref{sec:noiseless_case} we focus first on the noiseless case. The noisy case is then considered in Section~\ref{sec:noisy_case}, which also contains a comparison with separation-based schemes. Section~\ref{sec:conclusion} concludes the paper.
%
%
\subsection{Notational Remarks} \label{sec:notation}%
%
If multiplied by a matrix, a random length-$n$ sequence $X^n\coloneqq(X_1,\dots,X_n)$ is considered as a column vector. For $p\in[0,1]$, $H(p)= -p\log_2p-(1-p)\log_2(1-p)$ denotes the binary entropy function with the convention $0\log_20=0$. The Bernoulli distribution with parameter $p\in[0,1]$ is denoted as $\bern(p)$, which means that $X\sim\bern(p)$ takes on value $1$ with probability $p$. Addition modulo-2 is denoted as $\oplus$ and $\delta_{ij}$ represents the Kronecker delta, which is $1$ for $i=j$ and $0$ otherwise.
%
%
%
\section{System Model and Problem Statement}\label{sec:system_model}
Let $S_1,\dots,S_M$ be $M$ binary memoryless sources drawn from a joint probability mass function $P_{S_1\cdots S_M}$. In the presence of an eavesdropper, the sources are communicated to a legitimate receiver over a noisy channel. Unlike the usual setup in which the legitimate receiver wishes to reliably reconstruct each individual source while keeping the eavesdropper ignorant of them \cite{Tekin:Yener:08a,Ekrem:Ulukus:08,Goldenbaum:Schaefer:Poor:15}, in this paper the legitimate receiver is interested in reliably and securely computing a Boolean function
\begin{equation*}
	f:\{0,1\}^M\to\{0,1\}\;,\;U=f(S_1,\dots,S_M)
	\label{eq:desired_function}
\end{equation*}%
of the sources, to which we refer as the \emph{desired function}. 

\begin{figure}[!t]
	\centering
	\scalebox{1.2}{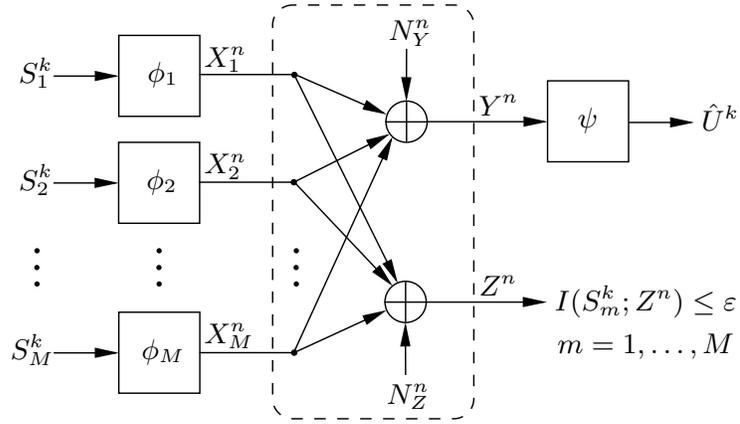}
	\caption{Secure computation over the binary modulo-2 adder multiple-access wiretap channel: a legitimate receiver wishes to reliably compute a function $U=f(S_1,\dots,S_M)$ of the sources while an eavesdropper has to be kept ignorant of them.}
	\label{fig:m2mawc}%
\end{figure}

In particular, as illustrated in Fig.\,\ref{fig:m2mawc}, we consider the toy scenario in which the channel between the sources and the destinations can be modeled as a memoryless \emph{binary modulo-2 adder multiple-access wiretap channel}, which is characterized by the input-output relations 
\begin{subequations}
	\begin{align}
		Y&=X_1\oplus\dots\oplus X_M\oplus N_Y\label{eq:legitimate}\;,\\
		Z&=X_1\oplus\dots\oplus X_M\oplus N_Z\label{eq:eavesdropper}\;.
	\end{align}%
	\label{eq:channel_outputs}%
\end{subequations}%
Here and hereafter, $X_m\in\{0,1\}$ is the channel input of source terminal $m$, $Y$ is the channel output seen by the legitimate receiver and $Z$ the output observed by the eavesdropper, respectively. The noise variables $N_Y\sim\mathrm{Bern}(p)$ and $N_Z\sim\mathrm{Bern}(q)$, for some $p,q\in[0,1/2]$, are assumed to be independent of the channel inputs.
\begin{remark}\label{rem:degraded}
	Note that each of the two multiple-access channels (MACs) in (\ref{eq:channel_outputs}) is a modulo-2 adder followed by a binary symmetric channel (BSC).
\end{remark}%

For some $k\in\mathds{N}$, $S_m^k$ denotes a length-$k$ sequence of independent and identically distributed samples of source $m$, $m=1,\dots,M$. In order to reliably compute at the legitimate receiver the sequence of corresponding function values, $U^k$, the source terminals employ a length-$n$ computation code defined as follows \cite{Nazer:Gastpar:07b}.
\begin{definition}\label{def:computation_code}
	Given a fixed desired function, a $(k,n)$ \emph{computation code} for the binary modulo-2 adder MAWC consists of the following:
	\begin{itemize}
		\item Encoding functions 
			\begin{equation*}
				\phi_m:\{0,1\}^k\to\{0,1\}^n\;,\;m=1,\dots,M\;,
				\label{eq:encoders}
			\end{equation*}%
			each of which maps $k$ source symbols to a length-$n$ codeword (i.e., $\phi_m(s_m^k)=x_m^n$);
		\item A decoding function at the legitimate receiver
			\begin{equation*}
				\psi:\{0,1\}^n\to\{0,1\}^k\;,
				\label{eq:decoder}
			\end{equation*}%
			 which maps each channel output sequence to a length-$k$ sequence of function values (i.e., $\psi(y^n)=\hat{u}^k$). 
	\end{itemize}%
\end{definition}%
The \emph{average probability or error} of a $(k,n)$ computation code is defined as
\begin{equation*}
	P_e^{(n)}\coloneqq\Prob\bigl[\hat{U}^k\neq U^k\bigr]\;,
	\label{eq:error_probability}%
\end{equation*}%
whereas the information about the source sequences \emph{leaked} to the eavesdropper is measured by
\begin{equation*}
	I(S_m^k;Z^n)\;,\; m=1,\dots,M\;,
\end{equation*}%
which we combine to the single constraint
\begin{equation}
	L^{(n)}\coloneqq I(S_1^k;Z^n)+\dots +I(S_M^k;Z^n)\;.
	\label{eq:leakage_rate}%
\end{equation}%
\begin{definition}\label{def:secrecy_computation_rate}
	For some given desired function, a rate $R\coloneqq k/n$ is said to be an \emph{achievable secrecy computation-rate} if there exists a sequence of $(nR,n)$ computation codes such that
	\begin{equation*}
		\lim_{n\to\infty}P_e^{(n)}=0\quad\text{and}\quad\lim_{n\to\infty}L^{(n)}=0\;.
	\end{equation*}%
\end{definition}%
\begin{definition}\label{def:secrecy_computation_capacity}
	For some given desired function, the \emph{secrecy computation-capacity} is defined as
	\begin{equation*}
		C_{\mathsf{sc}}\coloneqq\sup\{R:R\;\text{is an achievable secrecy computation-rate}\}.
	\end{equation*}%
\end{definition}%

Since the problem is challenging for arbitrary $f$, in this paper we focus on securely computing the modulo-2 sum of the source symbols: $f(s_1,\dots,s_M)=s_1\oplus\dots\oplus s_M$.
%
%
%
\section{The Noiseless Case}\label{sec:noiseless_case}%
First, in order to fix ideas and obtain insight, in this section we consider the noiseless case (i.e., $p=q=0$), which results in the channel outputs
\begin{equation*}
	Y = Z = X_1\oplus X_2\oplus\dots\oplus X_M\;.
	\label{eq:noiseless_adder}%
\end{equation*}%
For a certain class of joint source distributions, we have the following result. 
\begin{theorem}\label{thm:noiseless}
	 Let the desired function be the modulo-2 sum and the joint source distribution such that $P_{S_mU}=P_{S_m}P_U$ for all $m=1,\dots,M$. Then, the \emph{secrecy computation-capacity} is $C_{\mathsf{sc}}=1$ function values per channel use.
\end{theorem}%
\begin{IEEEproof} \emph{(Achievability).} Transmitting the source samples \emph{uncoded} results in the channel output sequences
	 \begin{equation*}
		 Z^k=Y^k=S_1^k\oplus\dots\oplus S_M^k=U^k
		 \label{eq:output_eve_noiseless}%
	 \end{equation*}%
	 and thus in $P_e^{(n)}\equiv 0$. On the other hand, we have   
	 \begin{equation*}
		 I(S_m^k;Z^k)=kI(S_m;Z)=kI(S_m;U)\;.
	 \end{equation*}%
	 But if $P_{S_mU}=P_{S_m}P_U$, then $S_m$ and $U$ are statistically independent and therefore $I(S_m;U)=0$. As this applies to all $m=1,\dots,M$, it follows for the leakage $L^{(n)}\equiv 0$. That is, we are able to reliably compute one function value per channel use while the eavesdropper is not able to obtain any information about the source sequences. 
	
	\emph{(Converse).} If we allow the encoders to fully cooperate, then the sum rate of the MAC in (\ref{eq:legitimate}) cannot exceed $\max_{P_{X_1\dots X_M}} I(X_1,\dots,X_M;Y)$, where $P_{X_1\dots X_M}$ denotes the joint distribution of the channel inputs. With or without secrecy constraint, we have
	\begin{align*}
		I(U;\hat{U})&\stackrel{\text{(a)}}{\leq}I(X_1,\dots,X_M;Y)\\
		&=H(Y)-H(Y|X_1,\dots,X_M)\\
		&=H(U)-H(U|S_1,\dots,S_M)\\
		&\stackrel{\text{(b)}}{=}H(U)\\
		&\leq 1\;,
	\end{align*}%
	which is a tight upper bound in our case. Note that (a) follows from the data processing inequality and (b) from the fact that $U$ is a function of $S_1,\dots,S_M$.
\end{IEEEproof}%

Due to the modulo-2 additivity of the channel along with the fact that the desired function perfectly matches this algebraic structure, the source sequences behave like \emph{one-time pads} protecting each other. Thus, the algebraic structure of the channel not only helps to efficiently compute the desired function at the legitimate receiver but also to protect the source sequences against eavesdropping. A remarkable fact is that the source terminals \emph{do not need} any additional source of randomness nor does the legitimate receiver need any advantage over the eavesdropper. This is in stark contrast to standard physical layer security problems in which a legitimate receiver typically wishes to securely decode messages. For instance, when the objective is to securely communicate messages over a MAWC, without local randomness the achievable secrecy rate region would be an empty set \cite{Tekin:Yener:08a,Ekrem:Ulukus:08,Goldenbaum:Schaefer:Poor:15}.   
\begin{remark}\label{rem:perfect_secrecy}
	Note that the coding strategy used in the proof of Theorem~\ref{thm:noiseless} achieves \emph{perfect secrecy}. Furthermore, the converse part of the proof implies that for the considered scenario, the secrecy computation-capacity equals the computation capacity $C_{\mathsf{c}}$. The latter is defined as the supremum over all achievable computation rates (i.e., without secrecy constraints) \cite{Nazer:Gastpar:07b}.
\end{remark}%

It is obvious that independent $\bern(1/2)$ sources fulfill the condition of Theorem~\ref{thm:noiseless} (i.e., $P_{S_mU}=P_{S_m}P_U$ for all $m=1,\dots,M$). Characterizing the set of all joint source distributions that fulfill this condition, however, is a nontrivial problem and beyond the scope of this paper. For the special case $M=2$, we have the following result.
\begin{theorem}\label{thm:condition}
	Let $U\coloneqq S_1\oplus S_2$. Then, $P_{S_mU}=P_{S_m}P_U$, $m=1,2$, if and only if $P_{S_1S_2}$ is doubly symmetric. That is, if and only if $P_{S_1S_2}$ is of the form 
	\begin{equation}
		P_{S_1S_2}(s_1,s_2)=\frac{1}{2}(1-\theta)\delta_{s_1s_2}+\frac{1}{2}\theta(1-\delta_{s_1s_2})\;,
		\label{eq:joint_source}%
	\end{equation}%
	for $\theta\in [0,1]$.
\end{theorem}%
\begin{IEEEproof}
	The proof is deferred to the Appendix.
\end{IEEEproof}%
%
%
%
\section{The Noisy Case}\label{sec:noisy_case}%
Now, we extend our considerations to the noisy case in which parameters $p$ and $q$ can be chosen arbitrarily (see (\ref{eq:channel_outputs})).
%
%
%
\subsection{Computation Capacity vs. Secrecy Computation-Capacity}\label{sec:comp_vs_seccomp}
%
Before presenting the main result of this paper, we recap a result that provides the computation capacity of the binary modulo-2 adder MAC given in (\ref{eq:legitimate}).
\begin{theorem}[Nazer\,-\,Gastpar \cite{Nazer:Gastpar:07b}]\label{thm:comp_cap}
	Let $f$ be the modulo-2 sum. Then, the \emph{computation capacity} of the binary modulo-2 adder MAC (\ref{eq:legitimate}) is given by
	\begin{equation*}
		C_{\mathsf{c}}=\frac{C}{H(U)}=\frac{1-H(p)}{H(U)}\;,
	\end{equation*}%
	where $C$ denotes the capacity of a BSC with crossover probability $p$.
\end{theorem}%

For the achievability part of the proof, Nazer and Gastpar employ random linear code ensembles for source compression and channel coding. By following their approach, we are able to extend Theorem~\ref{thm:noiseless} to the following. 
\begin{theorem}\label{thm:secrecy_comp_cap}
	Let $f$ be the modulo-2 sum and the joint source distribution such that $P_{S_mU}=P_{S_m}P_U$ for all $m=1,\dots,M$. Then, the \emph{secrecy computation-capacity} of the binary \mbox{modulo-2} adder MAWC is 
	\begin{equation*}
		C_{\mathsf{sc}}=C_{\mathsf{c}}=\frac{1-H(p)}{H(U)}\;.
	\end{equation*}%
\end{theorem}%
\begin{IEEEproof} \emph{(Achievability).}
	Let $C=1-H(p)$ denote the capacity of a BSC with crossover probability $p\in[0,1/2]$. 
	\begin{itemize}%
		\item \emph{Code construction:} Generate two matrices $A\in\{0,1\}^{n\times\ell}$ and $B\in\{0,1\}^{\ell\times k}$, each entry drawn uniformly and independently at random, with 
			\begin{equation}
				kH(U)<\ell <nC\;.
				\label{eq:l_cond}
			\end{equation}%
			Reveal $A$ and $B$ to the source terminals, the legitimate receiver, and the eavesdropper.
		\item \emph{Encoding:} Given $s_m^k$ at source terminal $m$, transmit 
			\begin{equation}
				x_m^n=\phi_m(s_m^k)=ABs_m^k\;,
				\label{eq:determ}%
			\end{equation}%
			where all operations are carried out modulo-2.
	\end{itemize}%

	With this encoding rule, the legitimate receiver observes the sequence of channel output symbols
	\begin{align}
		Y^n &= X_1^n\oplus\dots\oplus X_M^n\oplus N_Y^n\nonumber\\
		&=ABS_1^k\oplus\dots\oplus ABS_M^k\oplus N_Y^n\nonumber\\
		&=AB(\underbrace{S_1^k\oplus\dots\oplus S_M^k}_{=U^k})\oplus N_Y^n\label{eq:p2p}\;.
	\end{align}%
	Effectively, (\ref{eq:p2p}) is a BSC with crossover probability $p$. The random linear code induced by generator matrix $A$ therefore has the objective of protecting $BU^k$ against the noise $N_Y^n$, whereas the linear code induced by $B$ is used to compress $U^k$ to its entropy. As long as condition (\ref{eq:l_cond}) is fulfilled, there exist decoding functions $\psi':\{0,1\}^n\to\{0,1\}^{\ell}$ and $\psi'':\{0,1\}^{\ell}\to\{0,1\}^k$ such that for arbitrary $\varepsilon>0$ and $n$ large enough, the average probabilities of error (averaged over $A$ and $B$) fulfill $\Prob(\psi'(Y^n)\neq BU^k)<\frac{\varepsilon}{2}$ and $\Prob(\psi''(BU^k)\neq U^k)<\frac{\varepsilon}{2}$. This was shown in \cite{Nazer:Gastpar:07b} based on results from \cite{Elias:55} and \cite{Wyner:74}. Thus, defining the decoding function of Definition~\ref{def:computation_code} as 
	\begin{equation*}
		\psi(y^n)\coloneqq(\psi''\circ\psi')(y^n)\;,
	\end{equation*}%
	by means of the union of events bound we have $P_e^{(n)}<\varepsilon$ as long as $R=\frac{k}{n}<\frac{C}{H(U)}$ and $n$ sufficiently large.
	
	Now, we analyze the leakage. As in the proof of Theorem~\ref{thm:noiseless}, we consider each term of $L^{(n)}$ separately. Towards this end, 
	\begin{align*}
		I(S_m^k&;Z^n|A,B)\nonumber\\
		&=I(S_m^k;ABU^k\oplus N_Z^n|A,B)\nonumber\\
		&\leq I(S_m^k;ABU^k\oplus N_Z^n,U^k|A,B)\nonumber\\
		&=I(S_m^k;U^k|A,B)+I(S_m^k;ABU^k\oplus N_Z^n|A,B,U^k)\nonumber\\
		&=I(S_m^k;U^k)+I(S_m^k;N_Z^n|U^k)\nonumber\\
		&=0\;,
	\end{align*}
	where the last equality follows from the assumption $P_{S_mU}=P_{S_m}P_U$, the memorylessness of the sources, and the independence of $S_m$ and $N_Z$. As this applies to all $m=1,\dots,M$, we have $L^{(n)}\equiv 0$.
	
	\emph{(Converse).} For the average probability of error, $P_e^{(n)}$, to vanish with increasing block length, with or without a secrecy constraint every computation code has to fulfill
	\begin{align}
		kH(U)&\leq I(U^k;\hat{U}^k)\label{eq:first_line}\\
		&\stackrel{\text{(a)}}{\leq}I(X_1^n,\dots,X_M^n;Y^n)\nonumber\\
		&\leq\max_{P_{X_1,\dots,X_M}}I(X_1^n,\dots,X_M^n;Y^n)\nonumber\\
		&=n\bigl(1-H(p)\bigr)\label{eq:last_line}\;,
	\end{align}%
	where (a) is due to the data processing inequality. Combining the left hand side of (\ref{eq:first_line}) with (\ref{eq:last_line}) results in the upper bound $R=\frac{k}{n}\leq\frac{1-H(p)}{H(U)}$, which is tight in our case. 
\end{IEEEproof}%
\begin{remark}\label{rem:q_ind}
	It has to be emphasized that the secrecy com\-pu\-ta\-tion-capacity of Theorem~\ref{thm:secrecy_comp_cap} is independent of the MAC between the source terminals and the eavesdropper (i.e., independent of $q$). Note also that \emph{perfect secrecy} is achieved.
\end{remark}%

Surprisingly, the sequence of linear random codes that achieves the computation capacity also achieves the secrecy computation-capacity. No additional source of randomness is needed at the encoders as in the noisy case the source sequences act as one-time pads as well.
%
%
%
\subsection{Comparison with Separation-Based Computation}\label{sec:separation}
%
Consider the case $M=2$ and let $(S_1,S_2)$ be a doubly symmetric source with joint probability mass function given by (\ref{eq:joint_source}). By means of this explicit example, in this subsection we compare Theorem~\ref{thm:secrecy_comp_cap} with the secrecy computation-rate that is achievable with a separation-based coding scheme. A separation-based scheme first distributively compresses the source sequences into messages and then uses a capacity achieving MAC code in order to reliably reconstruct the messages at the legitimate receiver. Once $\hat{S}_1^k$ and $\hat{S}_2^k$ are known to the legitimate receiver it computes $\hat{U}^k=\hat{S}_1^k\oplus \hat{S}_2^k$, resulting in an estimate of the sequence of function values.

For this scenario, in \cite{Nazer:Gastpar:07b} it is shown that the best possible computation rate (i.e., without secrecy constraint) achievable with separation is
\begin{equation}
	R=\frac{1}{2}\left(\frac{1-H(p)}{H(\theta)}\right)\;.
	\label{eq:sep_comp_rate}%
\end{equation}%
The rate can be achieved with Körner-Marton compression for $U$ \cite{Koerner:Marton:79} in combination with time-sharing.\footnote{Note that for the two MACs given in (\ref{eq:channel_outputs}), time-sharing is optimal.} Compared with Theorem~\ref{thm:comp_cap}, this rate is only half the computation capacity. Because of time-sharing, however, when adding secrecy constraints the other source sequences may not act as one-time pads any longer so that local randomness has to be used at the encoders in order to confuse the eavesdropper.
\begin{theorem}\label{thm:separation}
	Let $M=2$, $f$ be the modulo-2 sum, and the joint source distribution as given in (\ref{eq:joint_source}). Furthermore, let $p\in[0,1/2)$ and $q= q'(1-2p)+p$ for some $q'\in (0,1/2]$. Then, for the binary modulo-2 adder MAWC, the best secrecy computation-rate achievable with separation is
	\begin{equation}
		R=\frac{1}{2}\left(\frac{H(q)-H(p)}{H(\theta)}\right)\;.
		\label{eq:sep_sec_rate}%
	\end{equation}%
	%
\end{theorem}%
\begin{IEEEproof} \emph{(Achievability).}
	As in the achievability part of the proof of Theorem~\ref{thm:secrecy_comp_cap}, the source terminals use the same linear random code for compressing $U$ to its entropy $H(U)=H(\theta)$. In \cite{Koerner:Marton:79}, Körner and Marton show that this is optimal for the joint source distribution given in (\ref{eq:joint_source}). Now, using time-sharing, the legitimate receiver alternately observes the channel outputs 
	\begin{equation}
		Y'=X_1\oplus N_Y\quad\text{and}\quad Y''=X_2\oplus N_Y
		\label{eq:bsc_leg}
	\end{equation}%
	while the eavesdropper sees
	\begin{equation}
		Z'=X_1\oplus N_Z\quad\text{and}\quad Z''=X_2\oplus N_Z\;.
		\label{eq:bsc_eve}
	\end{equation}%
	Thus, for each channel use we effectively have a binary symmetric wiretap channel of secrecy capacity
	\begin{align*}
		C(p)-C(q)&=1-H(p)-\bigl(1-H(q)\bigr)\\
		&=H(q)-H(p)\;,
	\end{align*}%
	where $C(q)$ denotes the capacity of the BSCs in (\ref{eq:bsc_leg}) and $C(p)$ the capacity of the BSCs in (\ref{eq:bsc_eve}), respectively.\footnote{For $p\in[0,1/2)$ and $q'\in (0,1/2]$, $q=q'(1-2p)+p>p$ and therefore $H(q)>H(p)$.} Thus, using standard wiretap coding allows $P_e^{(n)}$ and $L^{(n)}$ to be driven to zero as long as the sum rate fulfills
	\begin{equation*}
		k2H(\theta)< n\bigl(H(q)-H(p)\bigr)\;,
	\end{equation*}%
	which provides the rate in (\ref{eq:sep_sec_rate}).
	
	\emph{(Converse).} It can easily be checked that for $p\in[0,1/2)$ and $q'\in (0,1/2]$, the condition $q=q'(1-2p)+p$ implies an eavesdropper channel that is physically degraded with respect to the legitimate receiver's channel. In this case, the secrecy capacity region of the two-user binary modulo-2 adder MAWC is given by all rate pairs
	\begin{equation*}
		\bigl\{(R_1,R_2)\in\mathds{R}_+^2\,|\,R_1+R_2< H(q)-H(p)\bigr\}\;,
	\end{equation*}%
	which follows from \cite[Th.\,1]{Dai:Ma:14}. Thus, time-sharing in combination with single-user wiretap coding is optimal. 
\end{IEEEproof}%
%


After comparing (\ref{eq:sep_sec_rate}) with (\ref{eq:sep_comp_rate}), we conclude that se\-pa\-ration-based computation schemes generally suffer from imposing a secrecy constraint. In order to keep the source sequences secret from the eavesdropper, wiretap coding is needed and therefore local randomness at the encoders. This generally further reduces the achievable computation rate. 
\vspace{5pt}
%
%
\section{Conclusion} \label{sec:conclusion}
We have considered the problem of securely computing a function of distributed sources over the binary \mbox{modulo-2} adder MAWC. Instead of individual source samples, the legitimate receiver is interested in reliably decoding from the channel output a function of the sources. To characterize the corresponding fundamental limit, we have introduced the notion of secrecy computation-capacity and determined it for a function that perfectly matches the structure of the channel. Unlike standard results in physical-layer security, no additional randomness is needed in order to confuse the eavesdropper. 

Future work includes extensions to more general functions and MAWCs as well as to the case in which the joint source distribution does not fulfill the condition $P_{S_mU}=P_{S_m}P_U$, $m=1,\dots,M$. On the other hand, the leakage in (\ref{eq:leakage_rate}) might be replaced by another secrecy criterion such as $L^{(n)}=I(U^k;Z^n)$. This criterion is less restrictive as it prohibits the eavesdropper only from knowing anything about the function to be computed at the legitimate receiver.
%
%
%
\appendix[Proof of Theorem~\ref{thm:condition}] \label{app:proof}
Let $U\coloneqq S_1\oplus S_2$. We have to show that $P_{S_mU}=P_{S_m}P_U$, for $m=1,2$, if and only if the joint source distribution, $P_{S_1S_2}$, is doubly symmetric. That is, if and only if both of the equalities
\begin{subequations}
	\begin{align}
		P_{S_1S_2}(0,0)&=P_{S_1S_2}(1,1)\;,\\
		P_{S_1S_2}(0,1)&=P_{S_1S_2}(1,0)
	\end{align}%
	\label{eq:doubly_results}%
\end{subequations}%
hold. As the ``$\Leftarrow$'' part is trivial, we treat the ``$\Rightarrow$'' part only.

Note that for $P_{S_mU}=P_{S_m}P_U$ to be true, the following set of equations has to be fulfilled: 
\begin{subequations}
	\begin{align}
		P_{U|S_m}(0|0)&=P_{U|S_m}(0|1)\label{eq:doubly1}\;,\\
		P_{U|S_m}(1|0)&=P_{U|S_m}(1|1)\label{eq:doubly2}\;,
	\end{align}%
	\label{eq:doubly_symmetric_cond}%
\end{subequations}%
for $m=1,2$. As $u=0$ if and only if $(s_1,s_2)=(0,0)$ or $(s_1,s_2)=(1,1)$ and $u=1$ if and only if $(s_1,s_2)=(0,1)$ or $(s_1,s_2)=(1,0)$, the conditions in (\ref{eq:doubly_symmetric_cond}) can equivalently be expressed as
\begin{subequations}
	\begin{align}
		\frac{P_{S_1S_2}(0,0)}{P_{S_1}(0)}&=\frac{P_{S_1S_2}(1,1)}{1-P_{S_1}(0)}\label{eq:equiv_doubly1}\;,\\
		\frac{P_{S_1S_2}(0,1)}{P_{S_1}(0)}&=\frac{P_{S_1S_2}(1,0)}{1-P_{S_1}(0)}\label{eq:equiv_doubly2}\;,\\
		\frac{P_{S_1S_2}(0,0)}{P_{S_2}(0)}&=\frac{P_{S_1S_2}(1,1)}{1-P_{S_2}(0)}\label{eq:equiv_doubly1}\;,\\
		\frac{P_{S_1S_2}(0,1)}{1-P_{S_2}(0)}&=\frac{P_{S_1S_2}(1,0)}{P_{S_2}(0)}\label{eq:equiv_doubly2}\;.
	\end{align}%
	\label{eq:doubly_symmetric_cond_equiv}%
\end{subequations}%
Solving this system of equations subject to the constraints
\begin{align*}
	P_{S_1}(s_1)&=P_{S_1S_2}(s_1,0)+P_{S_1S_2}(s_1,1)\;,\\
	P_{S_2}(s_2)&=P_{S_1S_2}(0,s_2)+P_{S_1S_2}(1,s_2)
\end{align*}%
results in
\begin{equation*}
	P_{S_1}(0)=P_{S_2}(1)=P_{S_2}(0)=P_{S_2}(1)=1/2
\end{equation*}%
and thus in $\bern(1/2)$ marginals. Inserting this into (\ref{eq:doubly_symmetric_cond_equiv}) provides (\ref{eq:doubly_results}), which concludes the proof.
%
%
%
%
\vspace{5pt}
\end{document}